\documentclass[11pt,twoside]{article}
\usepackage{asp2010}
\usepackage{graphicx}

\resetcounters

\newcommand\an{{\em Astron.\ Nachr.}}

\newcommand{\Dnu}{\mbox{$\Delta \nu$}}
\newcommand{\bhyi}{\mbox{$\beta$~Hyi}}
\newcommand{\eboo}{\mbox{$\eta$~Boo}}
\newcommand{\muHz}{\mbox{$\mu$Hz}}

\newcommand{\Msun}{\mbox{$M_{\odot}$}}

\newcommand{\corot}{{\em CoRoT\/}}
\newcommand{\kepler}{{\em Kepler\/}}

\begin{document}

\title{Replicated \'Echelle Diagrams in Asteroseismology: A Tool for
  Studying Mixed Modes and Avoided Crossings}

\author{Timothy R. Bedding \affil{Sydney Institute for Astronomy (SIfA),
  School of Physics, University of Sydney, NSW 2006, Australia}}

\begin{abstract}
In oscillating stars, mixed modes have p-mode character in the envelope and
g-mode character in the core.  They are observed in subgiants and red
giants, in which the large density gradient outside the core effectively
divides the star into two coupled cavities.  This leads to the phenomenon
of mode bumping, in which mode frequencies are shifted from their regular
spacing as they undergo avoided crossings.  This short contribution
introduces a new diagram that involves extending the traditional \'echelle
diagram by replication.  This new diagram should prove a useful way of
displaying mixed modes and helping to identify missing modes that may lie
below the detection threshold.
\end{abstract}

\section{Mixed modes and avoided crossings} \label{sec.intro}

Asteroseismology involves using the oscillation frequencies of stars to
extract information about their parameters and internal structures.  For
solar-type main-sequence stars, the observed oscillations are pure p~modes
(acoustic waves, for which the restoring force arises from the pressure
gradient).  These are regularly spaced in frequency, closely following the
so-called asymptotic relation \citep{Tas80,Gou86}.  However, the
oscillations of post-main-sequence stars show departures from this
regularity that are due to the presence of mixed modes.

Mixed modes have p-mode character in the stellar envelope and g-mode
character in the core (g~modes are gravity modes, for which the restoring
force is buoyancy).  Mixed modes occur in evolved stars (subgiants and red
giants), in which the large density gradient outside the core effectively
divides the star into two coupled cavities.  This leads to the phenomenon
of {\em mode bumping}, in which mode frequencies are shifted from their
regular spacing
(for more details, see
\citealt{Scu74,Osa75,ASW77,D+P91,ChD2004,MME2008,AChDK2010,%
deheuvels+michel2010-crossings,deheuvels+michel2011-hd49385,bedding2011-canary}).

\begin{figure}

\centerline{
\includegraphics[angle=90,height=0.25\textheight]{fig.F3-un.epsi}
\includegraphics[angle=90,height=0.25\textheight]{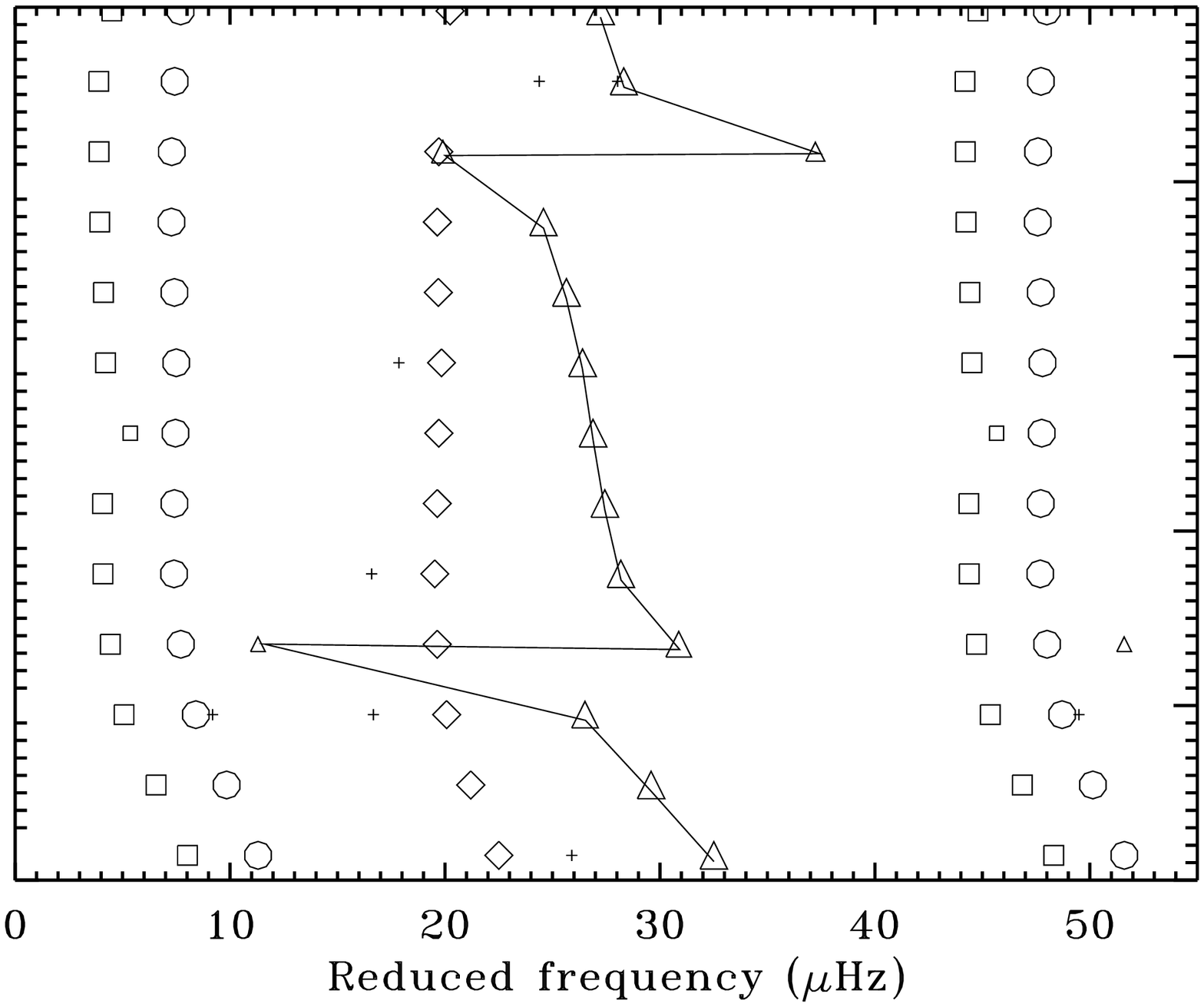}}

\caption{\label{fig.etaboo} Left: Evolution of oscillation frequencies in
models of a subgiant star of mass $1.60 \Msun$ and $Z = 0.03$ (representing
\eboo).  The dashed lines correspond to modes of degree $l = 0$, and the
solid lines to $l = 1$.  The vertical solid line indicates the location of
the model whose frequencies are illustrated in the right panel.  Right:
\'Echelle diagram using a frequency separation of $\Delta\nu = 40.3 \muHz$
and a zero-point frequency of $\nu_0 = 735 \muHz$.  Circles are used for
modes with $l = 0$, triangles for $l = 1$, squares for $l = 2$ and diamonds
for $l = 3$.  The size of the symbols indicates the relative amplitude of
the modes, estimated from the mode inertia, with crosses used for modes
whose symbols would otherwise be too small.  Figures adapted from
\citet{ChDBK95}. }
\end{figure}

Figure~\ref{fig.etaboo} shows theoretical oscillations frequencies for the
subgiant star \eboo, calculated by \citet{ChDBK95}.  The left panel shows
the evolution with time of the model frequencies for modes with $l=0$
(dashed lines) and $l=1$ (solid lines).  The $l=0$ frequencies decrease
slowly with time as the star expands.  At a given moment in time, such as
that marked by the vertical line, the radial modes are regularly spaced in
frequency, with a large separation of $\Dnu \approx 40\,\muHz$.  However,
the behaviour is rather different for the $l=1$ modes (solid lines).  They
undergo a series of {\em avoided crossings\/} as a function of time
\citep{Osa75,ASW77}, during which an $l=1$ mode is bumped upwards by the
mode below, and in turn it bumps the mode above.  The effect is to disturb
the regular spacing of the $l=1$ modes in the vicinity of each avoided
crossing, squeezing the modes together.  As discussed by
\citet{bedding2011-canary}, the frequencies of the avoided crossings
correspond to g~modes trapped in the core of the star (referred to as
$\gamma$~modes, following \citealt{ASW77}).

The right panel of Fig.~\ref{fig.etaboo} shows the frequencies in \'echelle
format \citep{GFP83} for the single model that is marked in the left panel
by the vertical line.  The bumping from regularity of the $l=1$ modes
(triangles) is obvious.  It is important to note there this is one extra
$l=1$ mode at each avoided crossing.

Bumping of $l=1$ modes has been observed and modelled in several
subgiant stars:
\begin{itemize}

\item \eboo\ \citep{KBV95,ChDBK95,G+D96,DiMChDK2003,KBB2003,Gue2004,CEB2005},
\item \bhyi\ \citep{DiMChDP2003,F+M2003,BKA2007,brandao++2011-bhyi}, 

\item the \corot\ target HD~49385
\citep{DBM2010,deheuvels+michel2010-crossings,deheuvels+michel2010-mu-gradient,deheuvels+michel2011-hd49385} 

\item the \kepler\ target KIC~11026764  (`Gemma'; \citealt{kepler-chaplin++2010-solar-like3,kepler-metcalfe++2010-gemma}),

\item the \kepler\ target KIC~11234888 (`Tigger'; \citealt{kepler-mathur++2011-boogie-tigger}),

\item the \kepler\ target KIC~11395018 (`Boogie'; \citealt{kepler-mathur++2011-boogie-tigger}),

\item the \kepler\ target KIC~10273246 (`Mulder'; \citealt{kepler-campante++2011-mulder-scully}) and

\item the \kepler\ target KIC~10920273 (`Scully'; \citealt{kepler-campante++2011-mulder-scully}).

\end{itemize}
For Procyon, which is close to the end of the main sequence,
\citet{BKC2010} suggested a possible $l=1$ mixed mode at low frequency,
based on the narrowness of the peak in the power spectrum.  Note that mixed
modes are expected to have longer lifetimes (smaller linewidths) than pure
p~modes because they have larger mode inertias (e.g., \citealt{ChD2004}).


The aim of this short contribution is to introduce a way of
extending the \'echelle diagram by replication, which may be useful in the
study of mixed modes.  Before doing so, we note that there are two ways to
make an \'echelle diagram.  One is to keep the orders horizontal --- the
\'echelle diagram in Fig.~\ref{fig.etaboo} uses this convention.
Alternatively, one can plot $\nu$ versus ($\nu \bmod \Dnu$), in which case
each order slopes upwards.  An example of this is discussed below
(Fig.~\ref{fig.rep-gemma}).

\section{The Replicated \'Echelle Diagram}

\begin{figure}
\centerline{
\includegraphics[width=\textwidth]{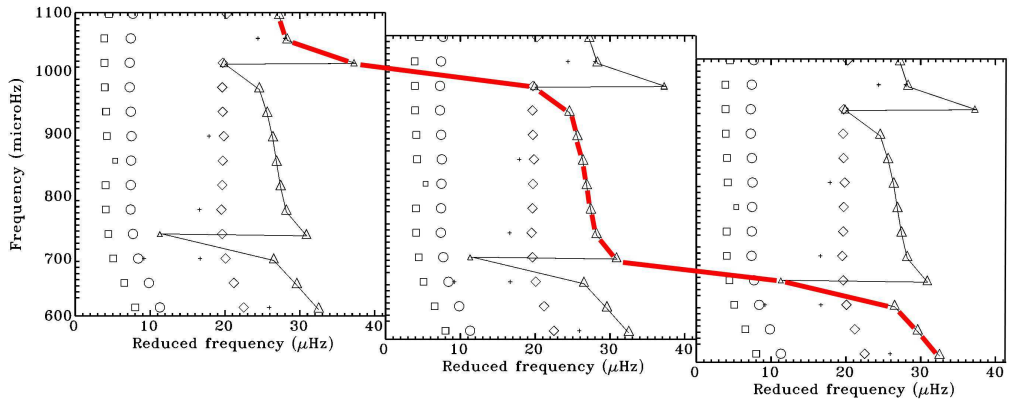}} 
\caption{\label{fig.rep-etaboo} Replicated \'echelle diagram of the model
of \eboo\ that is shown in the right panel of Fig.~\ref{fig.etaboo}.  The
thick red lines connect the $l=1$ modes. }
\end{figure}

Figure~\ref{fig.rep-etaboo} was made by replicating the \'echelle diagram
shown in Fig.~\ref{fig.etaboo}.  For reasons that will become clear, each
copy is displaced downwards by one order, that is, by \Dnu\ in frequency.
The thick red lines connect the $l=1$ modes and traces a smooth curve,
without the ``wrapping'' that occurs in a standard \'echelle diagram.

\begin{figure}
\centerline{
\includegraphics[width=\textwidth]{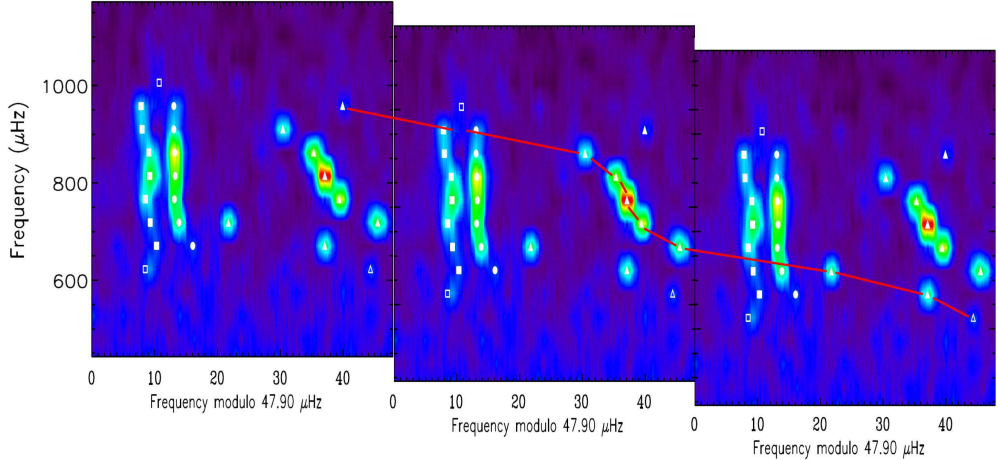}} 
\caption{\label{fig.rep-boogie} Replicated \'echelle diagram based on
  observations of KIC~11395018 (`Boogie') by
  \citet{kepler-mathur++2011-boogie-tigger}.  The red lines connect the
  $l=1$ modes.  }
\end{figure}

This type of figure allows us to see more clearly when modes are missing.
For example, Fig.~\ref{fig.rep-boogie} shows the replicated \'echelle
diagram for the \kepler\ target KIC~11395018 (`Boogie'), based on the power
spectrum observed by \citet{kepler-mathur++2011-boogie-tigger}.  As
mentioned above, we expect to see one extra $l=1$ mode at each avoided
crossing.  Because of the way the replicated \'echelle diagram is
constructed, with an offset of one order between each panel, we expect one
$l=1$ mode in every horizontal row.  In Fig.~\ref{fig.rep-boogie} we see
there is a missing $l=1$ mode near the top, which must lie between the
$l=0$ and $l=2$ modes (see the middle panel) and could be detected when
more data become available.

\begin{figure}
\centerline{
\includegraphics[width=\textwidth]{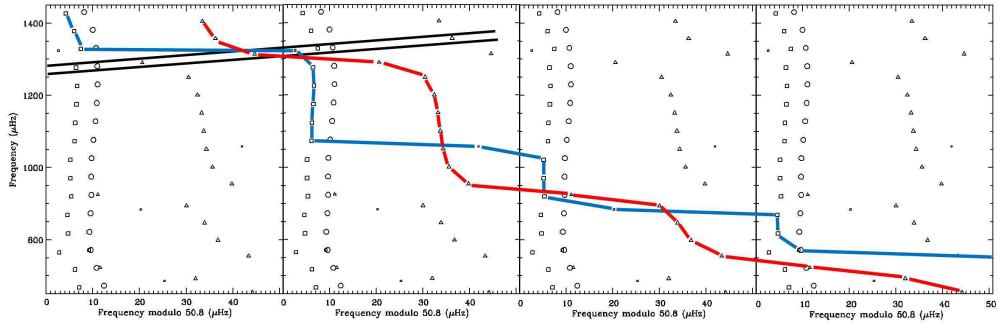}} 
\caption{\label{fig.rep-gemma} Replicated \'echelle diagram based on a
  model of KIC~11026764 (`Gemma'; Model~AA by
  \citealt{kepler-metcalfe++2010-gemma}).  Red and blue lines connect $l=1$
  and $l=2$ modes, respectively.  The black lines mark one of the
  upward-sloping orders.  }
\end{figure}

Another example is shown in Fig.~\ref{fig.rep-gemma}, using model
frequencies calculated by \citet{kepler-metcalfe++2010-gemma} for the
\kepler\ target KIC~11026764 (`Gemma').  This time we show two curves,
corresponding to $l=1$ modes (red) and $l=2$ modes (blue).  Note that each
curve moves to a new panel at each avoided crossing, and we see that the
avoided crossings for $l=2$ are more closely spaced than for $l=1$.  This
occurs because the frequencies of the avoided crossings correspond to
g~modes trapped in the core of the star (the so-called $\gamma$~modes ---
see above) and we know from asymptotic theory \citep{Tas80} that their
period spacing is inversely proportional to $\sqrt{l(l+1)}$ (see Eq.~3.236
of \citealt{AChDK2010}, where the quantity $L = \sqrt{l(l+1)}$ is defined
after Eq.~3.197).

The shape of each curve contains information about the strength of the
coupling between the p- and g-mode cavities.  As discussed by
\citet{deheuvels+michel2010-crossings,deheuvels+michel2011-hd49385}, the
coupling strength determines the details of the mode bumping, including the
number of modes that are affected.  In the replicated \'echelle diagrams,
the coupling strength could be measured at each avoided crossing from the
gradient of the inflection point.  We can see from Fig.~\ref{fig.rep-gemma}
that the coupling for the $l=2$ modes is much weaker than for $l=1$ modes,
as discussed by \citet{ChD2004}.

The replication in Fig.~\ref{fig.rep-gemma} does not include a vertical
offset between panels.  This is because this \'echelle diagram was made
with the sloping orders (one of which is enclosed by a pair of black
lines), whereas those in Figs.~\ref{fig.rep-etaboo}
and~\ref{fig.rep-boogie} were plotted with horizontal orders (see
Sec.~\ref{sec.intro}).  In both types of diagram, the construction method
ensures that the curve connects modes such that there is exactly one in
each `row' (whether the rows be horizontal or sloping).  Put another way,
the modes occur along the curve at the points where it intersects with each
row.  One can envisage fitting an analytical function to this curve; the
$\arctan$ function would appear to be a good choice.


\acknowledgements Many thanks to our Japanese hosts for a fantastic
workshop under very difficult conditions.  I also thank Othman Benomar,
J{\o}rgen Christensen-Dalsgaard, Hans Kjeldsen, Dennis Stello and Tim White
for helpful discussions.

\end{document}